\documentclass[
aps,
prl,
showpacs,
showkeys,
superscriptaddress,
groupedaddress,
]
{revtex4}
\usepackage{graphicx}  
\usepackage{dcolumn}   
\usepackage{bm}        
\usepackage{amssymb}   
\usepackage{multirow}
\usepackage[utf8]{inputenc}
\usepackage[english]{babel}
 
\usepackage{hyperref}
\hypersetup{
    colorlinks=true,
    linkcolor=blue,
    filecolor=magenta,      
    urlcolor=cyan,
}
 
\begin{document}

\title{Stability Analysis of an Interacting Holographic Dark Energy model}
\author{Sudip Mishra}
\email[]{sudipcmiiitmath@gmail.com}
\author{Subenoy Chackraborty}
\email[]{schackraborty.math@gmail.com}

\affiliation{Department of Mathematics, Jadavpur University, Kolkata- 700032, WB, India.}
\date{\today}

\begin{abstract}		
The present work deals with dynamical system analysis of Holographic Dark Energy cosmological model with different {\it infra-red} ({\it IR})-cutoff.  By suitable transformation of variables the Einstein field equations are converted to an autonomous system.  The critical points are determined and the stability of the equilibrium points are examined by Center Manifold Theory and Liapunov function method.  Possible bifurcation scenarios have also been explained.
\end{abstract}



\maketitle

\section{Introduction}	
	The evidences that our Universe is experiencing a phase of expansion with accelerated rate \cite{astro-ph/9812133}\cite{astro-ph/0402512} have been well demonstrated by a series of cosmological observations particularly from Type Ia Supernova, Cosmic Microwave Background Radiation (CMBR) anisotropies \cite{10.1086/513700}, Baryon Acoustic Oscillation \cite{astro-ph/0302209}, X-ray experiments and Large Scale Structures(LSS) \cite{10.1038/35010035}-\cite{astro-ph/9805201}.  This observational fact is accommodated in standard cosmology by introducing an unknown exotic matter known as dark energy having large negative pressure.  However, the nature of dark energy is still unrevealed, so that its nature is one of the major challenges in cosmology today.  The simplest model of dark energy is the cosmological constant $\Lambda$ which represents a vacuum energy density having equation of state $\omega_d=\frac{p_d}{\rho_d}=-1$ (where $p_d$ and $\rho_d$ are the thermodynamic pressure and energy density of the dark energy respectively).  This earliest and simplest theoretical candidate $(\Lambda CDM)$ was suggested in order to give a plausible explanation to the observational evidences (Universe's present day accelerated expansion) in a fair way.  It is well known that there are two main problems associated with $\Lambda$, namely the fine tuning and the cosmic coincidence problems.  In the former problem we see that the theoretical value of $\Lambda$ has many orders of magnitude larger than the current observational value (about an order of $10^{123}$ higher than what we observe\cite{1511.00600[gr-qc]})  while the later problem is related to unsolved mystery of densities of dark energy (DE) and dark matter (DM) of the same order at present although they evolve in different manner.  To solve these problems different dynamic dark energy  models \cite{1111.6247[gr-qc]}-\cite{1705.09908[astro-ph.CO]} have been suggested, with varying equation of state during the expansion of the Universe.  Among these Holographic Dark Energy (HDE) models have got much attention \cite{hep-th/0403052}\cite{10.1016/j.physletb.2004.10.014}.  HDE models are constructed based on the holographic principle, in quantum gravity, which states that the entropy of a system scales not with its volume but with its surface area $L^2$.  According to this principle, the zero point energy of the system with size L can be bounded by the mass of a blackhole with the same size \cite{hep-th/9803132} as follows
	\begin{equation}\label{eqn:1}
	\rho_{\Lambda}\leqslant L^{-2}M_p^2,
	\end{equation}     
	where $\rho_\Lambda$ is the quantum vacuum energy density and $M_p=\frac{1}{\sqrt{8\pi G}}$ is the plank mass.\\
	This inequality describes an analogy between the ultraviolet(UV) cut-off, defined through $\rho_\Lambda$ and the infra-red ({\it IR}) cut-off encoded in the scale L.  
	One can consider, in the context of cosmology, the dark energy density of the Universe $\rho_d$ to be same as the vacuum energy, i.e. $\rho_d=\rho_\Lambda$.  From effective quantum field theory, the largest {\it IR}-cutoff L is chosen by saturating the inequality.  As a result, the dark energy density (the vacuum energy density) can be written as \cite{hep-th/0403052, 1511.07955v1[gr-qc]}
	\begin{equation}\label{eqn:2}
	\rho_d=\frac{3M_p^2C^2}{L^2},
	\end{equation} 
	where ` C ' is a dimensionless numerical parameter which is estimated by observational data: for flat Universe (i.e. for k=0) it is obtained that $C=0.1818_{-0.097}^{+0.173}$ and in the case of non-flat Universe (i.e. for $k=\pm 1$) it is obtained that $C=0.815_{-0.139}^{+0.179}$ (reference \cite{0910.3855[astro-ph.CO]}\cite{0904.0928v2[astro-ph.CO]}).\\
	There are different choices of {\it IR}-cutoff of which three are widely used in the Literature, namely, Hubble radius\cite{10.1016/j.physletb.2004.10.014}\cite{hep-th/0403052}, Future event horizon\cite{10.1103/PhysRevLett.82.4971}\cite{FutureEventHorizon2}\cite{1005.3403[gr-qc]} and Ricci's scalar curvature\cite{0712.1394[astro-ph]}\cite{1101.4797[gr-qc]}.  The Hubble radius $L=H^{-1}$ ($H$ is the usual Hubble parameter) can not give the correct equation of state for dark energy but gives a correct energy density.  For the future event horizon $L=R_E$ and it is suggested that this choice of {\it IR}-cutoff may explain both the problems of the cosmological constant.  On the otherhand the choice $L=(\dot{H}+2H^2)^{-\frac{1}{2}}$ (the Ricci scalar curvature), introduced by Granda and Oliveros \cite{0810.3663[gr-qc]} is based on the spacetime scalar curvature as {\it IR}-cutoff and has its similarity with the size of maximal perturbation which leads to formation of blackhole.  Also this {\it IR}-cutoff may eliminate both the fine tuning and coincidence problems and is fairly good in fitting with the observational data.\\	
On the other hand, the interaction in the dark sector is a promising approach to address several cosmological problems existing for a long time. The appearance of the interaction was motivated to explain the tiny value of the cosmological constant \cite{hep-th/9408025}. Consequently,  it was realized that an interaction between dark matter and dark energy in a non-gravitational way could solve the cosmic coincidence problem \cite{astro-ph/9908023},  a severe problem existing in almost all non-interacting cosmological models. As a result, a considerable attention was paid in this field that resulted in  a series of interesting consequences \cite{astro-ph/9908224,astro-ph/0303228,astro-ph/0105479,gr-qc/0505020,0801.4233[astro-ph], 0804.0232[astro-ph],0806.2116[astro-ph],0812.2210[gr-qc]}. According to the astronomical data from various observational sources, recently, it has been reported in a series of articles that, interaction in the dark sector cannot be excluded although the coupling parameter characterizing the strength of the interaction is small enough \cite{1303.0684[astro-ph.CO],1401.1286[astro-ph.CO],1605.01712[astro-ph.CO],1701.00780[gr-qc],1702.02143[astro-ph.CO],1704.08342[astro-ph.CO],1706.04953[astro-ph.CO],1808.01669[gr-qc]} (also see \cite{1310.0085[astro-ph.CO]}). In addition to that, we would also like to focus that an interaction in the dark sector can solve the $H_0$ tension, see \cite{1702.02143[astro-ph.CO],1704.08342[astro-ph.CO]} where an extra degree of freedom  in terms of the coupling parameter plays the essential role. Thus, the interaction in the dark sector is a promising field of research for further investigations.\\
The holographic dark energy models interacting with dark matter can  solve the cosmic coindeince problem as well \cite{gr-qc/0505020, 0806.2116[astro-ph]} in contrary to the corresponding non-interacting cases. In addition to that, it has been also explored in some earlier works that if the future event horizon is chosen to be the {\it IR}-cutoff, then the equation of state for the holographic dark energy may cross the phantom divide line for some specific interaction models \cite{hep-th/0506069}. Such a scenario is also true in presence of the curvature of the Universe \cite{hep-th/0509107}. The crossing of phantom divide line is not possible for non-interacting holographic dark energy models, at least so far we are familiar with the literature. For a comparison of interacting and non-interacting holographic dark energy models, we refer to Ref. \cite{1612.00345[astro-ph.CO]}.\\  
The system of cosmological equations are nonlinear differential equations.  There is no well known method to find the exact solution of the system.  Dynamical system analysis is an elegant tool to study nonlinear systems.  Dynamical systems analysis allows us to gain a quantitative understanding of any cosmological model.  Apparently many recent research works are going on this approach to understand cosmological models both geometrical and physical point of view.  To survey some of the current researches we first cite the paper \cite{1604.07636[gr-qc]} where authors have taken interacting dark energy model (in the framework of particle creation mechanism).  Dynamical system analysis of this cosmological model in the background of spatial flat FLRW spacetime assembles some interesting cosmological scenarios.  The work of A. Paliathanasis \cite{1806.04969[gr-qc]} manoeuvres some advanced tools of  dynamical system to discuss cosmological viability of varying G(t) and $\Lambda(t)$ cosmology.  N. Ray {\sl et al.} \cite{1702.02169[gr-qc]},\cite{1708.07716[gr-qc]} analyze various cosmological models using dynamical system analysis.  S. Bahamonde {\sl et al.} \cite{1712.03107[gr-qc]} recently have published a review report to enhance an overview on the application of dynamical system in cosmology.  They have discussed various topics of nonlinear dynamics viz. linear stability theory, center manifold, Liapunov function, Poincar{\'e} sphere and behavior at infinity etc.  In 2015, N. Mahata \cite{1511.07955v1[gr-qc]} studied Interacting Holographic Dark Energy model with great details in cosmological point of view and analyzed some hyperbolic critical points (CPs) by Hartman-Grobman theorem and in 2017, Hanif Golchin \cite{1605.05068[gr-qc]} analyzed Interacting HDE model by dynamical system analysis with different interaction terms.  In our paper we reconstruct the autonomous system \cite{1511.07955v1[gr-qc]} so that the system can be $C^\infty (\mathbb{R}^2)$ and analyze all the CPs by center manifold and Liapunov function.  To analyze the stability of non-hyperbolic CPs center manifold theory has gained much importance in recent times \cite{1810.03816v1[gr-qc],1811.08279[gr-qc],1808.05634[gr-qc],1511.07978[gr-qc]}.  Possible bifurcation scenarios are also discussed in our work.
For this, we first assume Holographic Dark Energy model with event horizon as {\it IR}-cutoff.  Then we consider modified holographic dark energy model at Ricci's Scale (MHRDE).  This paper is organized as follows:  in the following sections, we discuss the basic equations and interaction terms of the model followed by the dynamical system analysis of interacting HDE model by transferring the field equations to autonomous systems and then interacting MHRDE model has been considered. At the end, cosmological implication of the dynamical system analysis has been discussed. 
	
\section{Basic Equations}

We consider the appearance of the Universe to be homogeneous and isotropic flat FRW spacetime and assume that it is interacting with DM in the form of dust (having energy density $\rho_m$) and HDE in the form of perfect fluid having variable equation of state $\omega_d=\frac{p_d}{\rho_d}$ where $p_d$ and $\rho_d$ are the thermodynamic pressure and energy density of the dark energy respectively.  \par

The Einstein field equations for spatially flat model are,
\begin{equation} \label{eqn:3}
3H^2=\rho_m + \rho_d
\end{equation}
and
\begin{equation}\label{eqn:4}
2\dot{H}=-\rho_m-(1+\omega_d)\rho_d,
\end{equation}
with $8\pi G=1$ for simplicity.\\
Also, the energy conservation equations for the fluids can be expressed as
\begin{equation}\label{eqn:9}
\dot{\rho_m}+3H\rho_m = Q
\end{equation}
and
\begin{equation}\label{eqn:10}
\dot{\rho_d}+ 3H(1+\omega_d)\rho_d=-Q,
\end{equation}
where `{\bf $\cdot$}' represents the derivative with respect to time  `{\bf t}'.  The interaction term Q is not unique. It is to be noted that energy is transferred from DE to DM for $Q>0$.  Usually the positivity of Q is justified for the following reasons
\begin{itemize}
\item It ensures the validity of the second law of thermodynamics.
\item It satisfies the Le Chatelier's principle\cite{0712.0565[gr-qc]}.
\item A possible solution of the coincidence problem.
\end{itemize}
Baryonic matter is not included in the interaction due to the constraints which are measured by local gravity. \par

Unfortunately, there is no definite theory yet to derive the interaction rate between the dark matter and dark energy since the nature of these dark fluids are absolutely unknown. 
Thus, usually some phenomenological models are considered to understand their effects on the underlying space-time. But the impacts of interaction are significant as explained earlier. Here we assume some mostly used and well known interaction models as follows. $Q = 3 H b^2 \rho$ \cite{0801.4233[astro-ph]}, $Q = 3 H \nu \rho_d $ \cite{1109.6234[astro-ph.CO]} and $Q = \frac{\nu}{H} \rho_m \rho_d$ \cite{0801.4233[astro-ph]}. All three models have been widely studied in the literature. In addition to that, for the first two interaction models, the evolution equations for DM and DE can be analytically solved. Since for all three interaction models, due to the presence of the coupling parameter (either $b^2$ or $\nu$), the corresponding parameters space is increased compared to the six-parameters based $\Lambda$CDM model, thus, the models could effectively take higher values of $H_0$ and thus, the tension on $H_0$ is released. This particular feature is not present for a general class of non-interacting $\Lambda$CDM type cosmologies.\par

The acceleration of the Universe can be derived using the field Eqns. (\ref{eqn:3}) and (\ref{eqn:4}) and is given by,
\begin{equation}
\ddot{a}=a(\dot{H}+H^2)=-\frac{a}{6}\lbrace \rho_m +(1+\omega_d)\rho_d \rbrace.
\end{equation}
It is to be noted that present accelerating phase corresponds to $\omega_d<-\frac{1}{3}$.\par

For the present model the deceleration parameter can be written as $q=\frac{1}{2}(1+3\omega_T)$, with $\omega_T=\frac{p_d}{\rho_m+\rho_d}=\omega_d\Omega_d$, where 
\begin{equation}
\Omega_d=\frac{\rho_d}{3H^2}
 \label{eqn:8}
\end{equation}
 is the density parameter for the dark energy and $w_T$ is the equation of state parameter for the combine single fluid.  
 
\section{Holographic DE Model with Event Horizon as {\it IR}-Cutoff}\label{sec:3}
The choice of the {\it IR}-cutoff as event horizon of the Universe ($R_E$) is well accepted \cite{1309.3136[astro-ph.CO]} that is the most suitable choice for the {\it IR}-cutoff where $R_E$ is defined by the improper integral \cite{1511.07955v1[gr-qc], hep-th/0403052}:

\begin{equation}\label{eqn:12}
R_E=a\int_{t}^{\infty} \frac{dt}{a}.
\end{equation}
It is to be noted that above improper integral converges only when strong energy condition is violated.  So in the present accelerating phase the integral always exists.  Now choosing $L=R_E$, we have from (\ref{eqn:2})
\begin{equation}\label{eqn:13}
\rho_d= \frac{3{M_p}^2C^2}{R_E^2}.
\end{equation}
Using this expression for the energy density $\rho_d$ in the conservation Eqn. (\ref{eqn:10}), the equation of state parameter is obtained (for any interaction) as  
\begin{equation}\label{eqn:14}
\omega_d=-\frac{1}{3}-\frac{2\sqrt{\Omega_d}}{3C}-\frac{Q}{3H\rho_d}.
\end{equation}

We now analyze the evolution equations for different choices of interaction term separately.

\begin{center}
A. $Q=3b^2H\rho=3b^2H(\rho_m+\rho_d)$
\end{center}
From Eqns. (\ref{eqn:4}), (\ref{eqn:9}) and (\ref{eqn:10}) we get an autonomous system with variables $H,\rho_m,\rho_d$.  But systems in the $(H,\rho_m)$ , $(H,\rho_d)$ and $(\rho_m,\rho_d)$ planes are equivalent in the sense that underlying variables are related by Eqn. (\ref{eqn:3}).  Now for the sake of simplicity of visualizing the phase space of bouncing cosmological solution we consider the autonomous system in $(H,\rho_d)$-plane.  We also analyze the system for the case $b=1$ which is not viable in $(\rho_m,\rho_d)$-plane.  We find the bifurcation values according to expansion and contraction of the Universe in the $(H,\rho_d)$-plane.  The stability analysis of autonomous system in $(H,\rho_m)$ gives similar physical information to the $(H,\rho_d)$-plane.  So we skip this case.\\
 Next we analyze the autonomous system in $(\rho_m,\rho_d)$-plane to find the bifurcation values according to stability of the system with respect to energy density ($\rho_m$) of DM and energy density ($\rho_d$) of DE.  Finally we restrict the vector fields on parabolic cylinder (satisfying Eqn.(\ref{eqn:3})) embedded in $\mathbb{R}^3$ and analyze non-static model of the Universe in $(H,\rho_m,\rho_d)$-space. \\      
The second Friedman equation (\ref{eqn:4}) and the energy conservation for DE (i.e. equation (\ref{eqn:10})) can be reduced (after a bit simplification) into an autonomous system parallel to $(H,\rho_d)$ plane as 
\begin{eqnarray}
\dot{H}&=& -\frac{3H^2}{2}[1-\frac{\Omega_d}{3}-\frac{2{\Omega_d}^\frac{3}{2}}{3C}-\frac{Q}{3H\rho}] \label{eqn:15}\\
\dot{\rho_d}&=& 2\rho_d H[\frac{\sqrt{\Omega_d}}{C}-1] \label{eqn:16}.
\end{eqnarray}

We choose the variable  $\rho_d= v^2$, the Eqns. (\ref{eqn:15}) and (\ref{eqn:16}) generate (using Eqns. (\ref{eqn:3}) and (\ref{eqn:8})) an autonomous system in (H,v)-plane as follows 
\begin{eqnarray}
\dot{H}&=&-\frac{1}{2}[3H^2(1-b^2)-\frac{v^2}{3}-\frac{2v^3}{3\sqrt{3}CH}] \label{eqn:17}\\ 
\dot{v}&=&v[\frac{v}{\sqrt{3}C}-H]. \label{eqn:18}
\end{eqnarray}
This is a continuously differentiable system on $\mathbb{R}^2$ for $H\neq 0$.  Henceforth, we tag subscript/superscript `c' with a variable as critical point of the corresponding system. The critical points of the system (\ref{eqn:17}-\ref{eqn:18}) are  $(H_c,\sqrt{3}CH_c)$ for all $H_c \in \mathbb{R}-\lbrace 0 \rbrace$, provided $1-b^2=C^2$.  For $b \neq 1$ the eigenvalues of the Jacobian matrix (at CPs) are $\lbrace(1-4c^2)H_c,0\rbrace $.  So the critical points are non-hyperbolic in nature.  The eigenvector corresponding to 0 is $(1~,~\sqrt{3}C)^T$ and the eigenvector corresponding to $(1-4C^2)H_c$ is $(1~,~\frac{\sqrt{3}}{4C})^T$.  The linear part of the system can be reduced to the Jordan form.  The corresponding Jordan matrix is the following
\[
J_{Jordan}(H_c,\sqrt{3}CH_c)=
\left[ {\begin{array}{cc}
	(1-4C^2)H_c & 0 \\
	0 & 0 \\
	\end{array} } \right].
\]
This yields the following system in diagonal form
\begin{eqnarray}
\dot{H}&=&[(1-4C^2)H_c]H + \mathcal{O}(H^2) \label{eqn:19}\\
\dot{v}&=& \mathcal{O}(v^2). \label{eqn:20}
\end{eqnarray}

\begin{figure}
\includegraphics[scale=0.45]{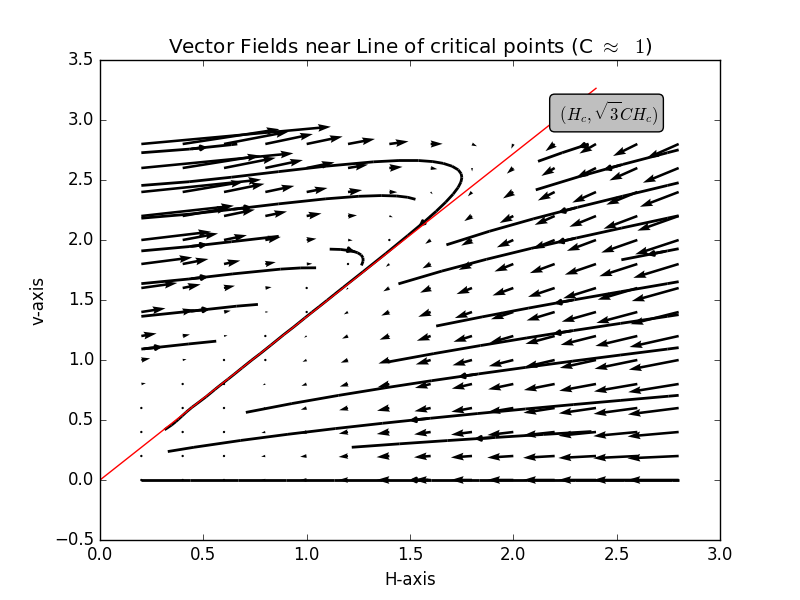}
\caption{\label{fig:1} For $H_c>0$ .}
\end{figure}

\begin{figure}
	\includegraphics[scale=0.45]{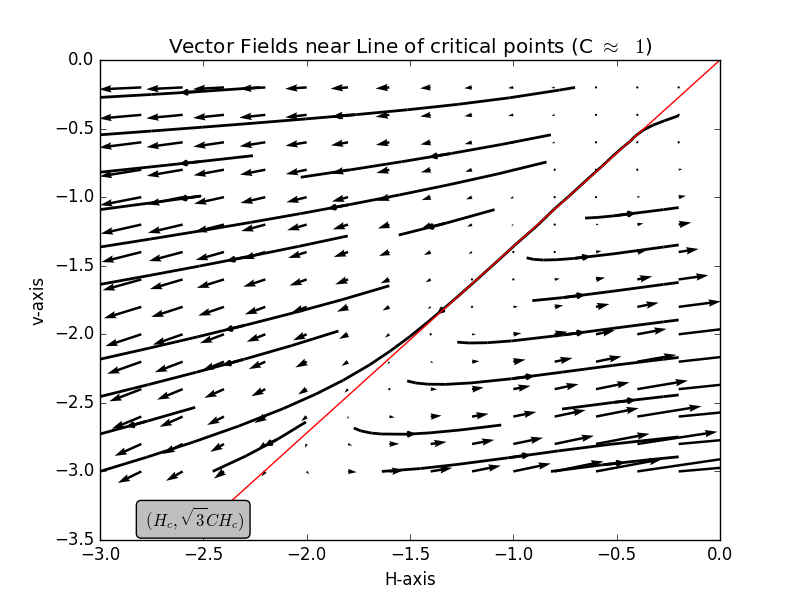}
	\caption{\label{fig:2} For $H_c<0$ .}
\end{figure}

The line of non-hyperbolic critical points $(H_c,\sqrt{3}CH_c)$ are normally hyperbolic \cite{978-94-017-0327-7, 1604.07636[gr-qc], 1811.08279[gr-qc]}. The stability of normally hyperbolic set can be completely classified by considering the sign of the eigenvalues in the remaining directions.  Due to Hartman-Grobman theorem (sec. 2.8 in \cite{978-1-4613-0003-8}) the flow is attracting towards the CP when $(1-4C^2)H_c<0$ and repelling when $(1-4C^2)H_c>0$ along the eigenvector $(1~,~\frac{\sqrt{3}}{4C})^T$ (see Figure \ref{fig:1} and \ref{fig:2}).\par
In this case, the trajectory can flow from $H_c >0$ to $H_c <0$ i.e. there is a bouncing cosmological solution from expanding phase to the contracting phase of the Universe. \par
At any critical point the system (\ref{eqn:17}-\ref{eqn:18}) is structurally unstable for $C=\pm \frac{1}{2}$.  So each non-hyperbolic critical point $(H_c, \sqrt{3}CH_c)$ is a bifurcation point at the bifurcation values $C= \pm \frac{1}{2}$.\par
The equation of state parameter $(\omega_d)$ at CPs $(H_c,\sqrt{3}CH_c)$ can be determined by $-\frac{1}{C^2}$ and the corresponding deceleration parameter $(q)$ is $-1$.\par 

For $b^2=1$ the system (\ref{eqn:17},\ref{eqn:18}) reduces to the following system 
\begin{eqnarray}
\dot{H}&=&\frac{v^2}{2}[\frac{1}{3}+\frac{2v}{3\sqrt{3}CH}] \label{eqn:21}\\
\dot{v}&=&v[\frac{v}{\sqrt{3}C}-H]. \label{eqn:22}
\end{eqnarray}
The critical points (CPs ) are $(H_c,0)$ where $H_c\in \mathbb{R}-\lbrace 0 \rbrace$.  The eigenvalues of Jacobian Matrix at $(H_c,0)$ are $\lbrace0,-H_c\rbrace$.  So all the critical points are normally hyperbolic in nature.  The vector fields along v-axis is attracting towards the CP for $H_c>0$ and repelling for $H_c<0$. 
The above analysis shows that the trajectory can flow from $H_c >0$ to $H_c <0$ i.e. there is a bouncing cosmological solution from expanding phase to the contracting phase.  This is also reflected in the Center Manifold (\cite{HASSARD1978297}, sec. 2.12 in \cite{978-1-4613-0003-8}) v=0  at the origin (after shifting the CP to origin).  Thus from (\ref{eqn:21}) we get $\dot{H}=0$ and one gets $q=\frac{1}{2}$.\\  
Further, for the interaction $Q=3b^2H\rho$, the conservation Eqns. (\ref{eqn:9}) and (\ref{eqn:10}) can be expressed in the form 
\begin{eqnarray}
\dot{\rho_m}&=&\sqrt{3(\rho_m+\rho_d)}[b^2\rho_d-(1-b^2)\rho_m] \label{eqn:23}\\
\dot{\rho_d}&=& - \sqrt{3(\rho_m+\rho_d)}[b^2\rho_m + (1+\omega_d+b^2)\rho_d]. \label{eqn:24}
\end{eqnarray}

The above system is well defined and continuously differentiable on the open set $E=\lbrace (\rho_m,\rho_d)\in \mathbb{R}^2 | \rho_m>0, \rho_d>0\rbrace \subset \mathbb{R}^2$.  The critical points are $(\rho_m^c,\rho_d^c)\in E$ where $\rho_m^c=\frac{b^2}{(1-b^2)}\rho_d^c$.  Here we consider $1-b^2=C^2$ so $\omega_d=\frac{1}{b^2-1}$ $(\leqslant -1)$ and $0\leqslant b^2<1$.  At this set of critical points $\Omega_d=1-b^2$, $\omega_T=-1$ and the deceleration parameter q=-1.\\
As all the critical points are non-hyperbolic critical points we find a suitable Liapunov function (sec 2.9 in \cite{978-1-4613-0003-8}) so that we can find the stability of the system for all critical points.\\
We consider $V(\rho_m,\rho_d)=(\rho_m-\frac{b^2}{1-b^2}\rho_d)^2$, which is a continuously differentiable function over $\mathbb{R}^2$ and $V(\rho_m^c,\rho_d^c)=0$ at the line of critical points $(\rho_m^c,\rho_d^c)$ and $V(\rho_m,\rho_d)>0$ for all other points.
\begin{widetext}
\begin{equation}\label{eqn:25}
\dot{V}(\rho_m,\rho_d)=2(\rho_m-\frac{b^2}{1-b^2}\rho_d)(\dot{\rho_m}-\frac{b^2}{1-b^2}\dot{\rho_d})=2 \frac{2b^2-1}{1-b^2}\sqrt{3\rho}(\rho_m-\frac{b^2}{1-b^2}\rho_d)^2.
\end{equation} 
\end{widetext}
Now it is to be noted that
\begin{itemize}
	\item for $b^2\in [0,\frac{1}{2})$, $\dot{V}(\rho_m,\rho_d)<0$ for all points in E except critical points.  This implies that all the critical points are asymptotically stable.
	\item for $b=\pm \frac{1}{\sqrt{2}}$, $\dot{V}(\rho_m,\rho_d)=0$ for all points in E.  So, all the critical points are stable. 
	\item for $b^2\in (\frac{1}{2},1)$, $\dot{V}(\rho_m,\rho_d)>0$ for all points in E except critical points.  This implies that all the critical points are unstable.
\end{itemize}
Thus the system (\ref{eqn:23}-\ref{eqn:24}) is structurally unstable at $b=\pm \frac{1}{\sqrt{2}}$.  So each non-hyperbolic critical point $(\rho_m^c,\rho_d^c)$ is a bifurcation point\cite{HASSARD1978297} at the bifurcation values $b=\pm \frac{1}{\sqrt{2}}$.\\
 To analyze a non-static model on hyperbolic cylinder we invoke 2D-autonomous system (\ref{eqn:23}-\ref{eqn:24}) together with Eqn.
\begin{equation}\label{eqn:24b}
\dot{H}=-\frac{1}{2}[3H^2-\frac{\rho_d}{1-b^2}].
\end{equation}
 The above system is well defined and continuously differentiable on the open set $\OE=\lbrace (H,\rho_m,\rho_d)\in \mathbb{R}^3 | H,\rho_m>0, \rho_d>0\rbrace \subset \mathbb{R}^3$ and we take the same restriction of b as we consider the previous 2D-system. \\
 In this case we choose the Liapunov function $V(H,\rho_m,\rho_d)=(\rho_m-\frac{b^2}{1-b^2}\rho_d)^2 + (3H^2-\frac{\rho_d}{1-b^2})^2$.
\begin{widetext}
\begin{equation}\label{eqn:25b}
\dot{V}(H,\rho_m,\rho_d) = 2 \frac{2b^2-1}{1-b^2}\sqrt{3\rho}[(\rho_m-\frac{b^2}{1-b^2}\rho_d)^2 +(3H^2-\frac{\rho_d}{1-b^2})^2].
\end{equation} 
\end{widetext}
 So the stability analysis of $V(H,\rho_m,\rho_d)$ is same as we have done for $V(\rho_m,\rho_d)$ with same set of bifurcation values but the trajectories recline on the parabolic cylinder (satisfying Eqn. (\ref{eqn:3}) and topologically equivalent to $E$) embedded in the $(H,\rho_m,\rho_d)$-space.


\begin{center}
B. $Q=\frac{\nu}{H}\rho_m\rho_d, (\nu>0)$
\end{center}
For this choice of interaction term and considering $\Omega_d=u^2$, the equation of state parameter (\ref{eqn:14}) takes the form 
\begin{equation}\label{eqn:26}
\omega_d=-\frac{1}{3}-\frac{2u}{3c}-\nu(1-u^2)
\end{equation}
and the evolution of density parameter is modified as
\begin{equation}\label{eqn:27}
\dot{u}=\frac{H}{2}u(1-u^2)[1-3\nu u^2 + \frac{2u}{C}].
\end{equation}
The second Friedman equation can be expressed in terms of density parameter as
\begin{equation}\label{eqn:28}
\dot{H}=-\frac{3H^2}{2}[1-\frac{u^2}{3}-\frac{2u^3}{3C}-\nu u^2(1-u^2)].
\end{equation}
The above system of Eqns. (\ref{eqn:27}) and (\ref{eqn:28}) form an autonomous system in (u,H)-plane.  The right hand side of the system (\ref{eqn:27}-\ref{eqn:28}) is a $C^\infty(\mathbb{R}^2)$ non-linear function for C being a non-zero constant.   The CPs of the above system are $(u_c,H_c)$ where $u_c,H_c ~\in~\mathbb{R}- \lbrace 0 \rbrace$ provided $\nu=\frac{1}{u_c^2}$ and $C=u_c$.  The eigenvalues of the Jacobian matrix at CPs are $\lbrace 2H_c(u_c^2-1),0\rbrace$.  As the critical points are normally hyperbolic, the stability of the CPs are determined by sign of the corresponding eigenvalue $2H_c(u_c^2-1)$.  In this case, $\omega_d=-\frac{1}{\Omega_d}$ and the deceleration parameter $q=-1$ (table \ref{table1}).\par
\begin{figure}[h]
	\includegraphics[scale=0.45]{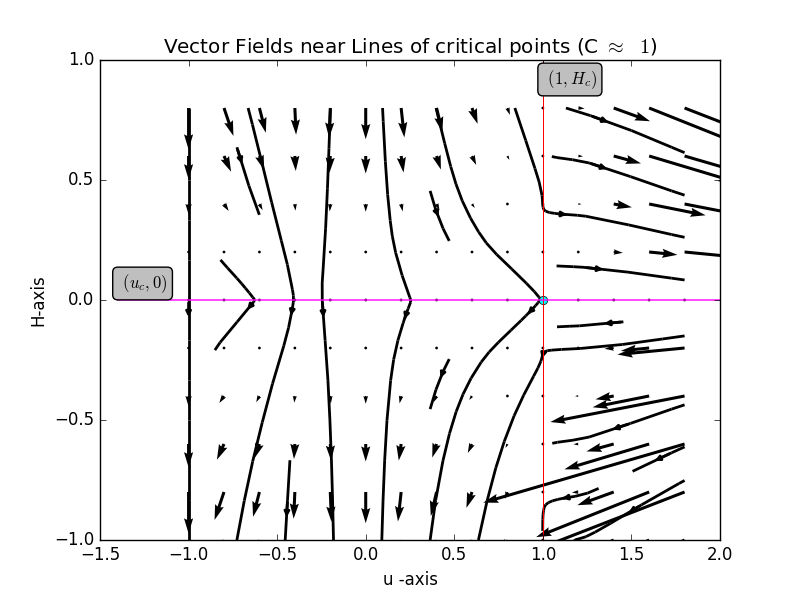}
	\caption{\label{fig:3} stability of line of critical points.}
\end{figure}

In view of $C=1$, $(1,H_c)$ are critical points for all $H_{c}\in \mathbb{R}$ and corresponding eigenvalues of Jacobian matrix are $\lbrace 0,0\rbrace$.  So stability of these CPs are not conclusive as there is no nonzero eigenvalue of the Jacobian matrix evaluated at the critical point. 
  For similar reason we do not able to analyze the stability of line of CPs $(u_c,0)$ where $u_c~\in~\mathbb{R}$ by dynamical system analysis (see Figure \ref{fig:3}). Moreover, the density parameter remains undefined for this set of CPs.\par

\begin{table}[ht]
	\caption{ Stability analysis}
	\begin{tabular}{ |c|c|c|c|c|c| }
		\hline
		critical Points  & $H_c$ & $\nu$ & $\omega_d$ & q & stability\\
		\hline
	    \multirow{5}{*}{$(u_c,H_c)$} & $>0$ & $> 1$ & -1 & -1 & stable\\ 
		
		& $<0$ & $> 1$ & -1 & -1 & unstable\\ 
		
		& $>0$ & $< 1$ & -1 & -1 & unstable\\ 
		
		& $<0$ & $<1$ & -1 & -1 & stable\\ 
		
		& $>$ or $<$ & = 1 & -1 & -1 & undetermined\\
		\hline
		
	\end{tabular}\label{table1}
\end{table}

From Table \ref{table1} we conclude that the system (\ref{eqn:27}-\ref{eqn:28}) is structurally unstable at $\nu=1$.  So $\nu=1$ is a bifurcation value at each normally hyperbolic CP $(1,H_c)$ which is dubbed bifurcation point.

\begin{center}
	C. $Q = 3\nu H\rho_d, ~(\nu>0)$
\end{center}

For this choice of interaction term and considering $\Omega_d=u^2$, the equation of state parameter (\ref{eqn:14}) takes the form 
\begin{equation}\label{eqn:29}
\omega_d=-\frac{1}{3}-\frac{2u}{3c}-\nu.
\end{equation}
The evolution of density parameter and the second Friedman equation can be expressed by an autonomous system in $(u,H)$ plane as follows
\begin{eqnarray}
\dot{u}&=&\frac{H}{2}[u+\frac{2}{C}u^2-u^3(1+3\nu)-\frac{2}{C}u^4] \label{eqn:30}\\
\dot{H}&=&-\frac{3}{2}H^2[1-\frac{u^2}{3}-\frac{2u^3}{3C}-\nu u^2]. \label{eqn:31}
\end{eqnarray}

For non-zero H and u the critical points of the system (\ref{eqn:30}-\ref{eqn:31}) are $(u_c,H_c)$ where $u_c,H_c ~\in~\mathbb{R}- \lbrace 0 \rbrace$ provided $C=u_c<1$ and $\nu=\frac{1-u_c^2}{u_c^2}$.  The eigenvalues of the Jacobian matrix at the CPs are $\lbrace H_c(7-10u_c^2),0\rbrace$.  This set of normally hyperbolic CPs can be analyzed by the sign of non zero eigenvalue(s).  For the above choices of C and $\nu$, we get $\omega_d=-\frac{1}{u_c^2}$ and decelerating parameter $q=-1$.  We get different critical points for different choices of C and $\nu$.  Moreover, all the critical points are non-hyperbolic as well as normally hyperbolic in nature.\\
For H=0, we get a line of CPs $(u_c,0)$ where $u_c \in \mathbb{R}$.  We do not get any nonzero eigenvalue of the Jacobian matrix at these CPs.  So nature of the CPs remains undetermined by dynamical system analysis.  Moreover, the density parameter remains undefined for this set of CPs.


\begin{center}
\textbf{Modified Holographic Ricci Dark Energy Model}
\end{center}
Here we express the modified holographic Ricci dark energy by taking the {\it IR}-cutoff with the modified Ricci radius in terms of $\dot{H}$ and $H^2$ as
\begin{equation}\label{eqn:32}
\rho_d=\frac{2}{\alpha-\beta}(\dot{H}+\frac{3\alpha}{2}H^2),
\end{equation} 
where $\alpha, ~\beta$ are free constants and take the value as mentioned in \cite{1309.3136[astro-ph.CO], astro-ph/0105479}.\\
The equation of state parameter for DE takes the form
\begin{equation}\label{eqn:33}
\omega_d= -(\alpha-\beta) + \frac{\alpha}{\Omega_d}-\frac{1}{\Omega_d}
\end{equation}
and the deceleration parameter is given by
\begin{equation}\label{q2}
q=\frac{3}{2}\alpha -1 -\frac{3}{2}\Omega_d(\alpha-\beta).
\end{equation}

We again analyze the evolution equations for the given choices of interaction term separately:\\
\begin{center}
A. $Q=3b^2H\rho$, ($0\leqslant b^2<1$)
\end{center}
Using field Eqn. (\ref{eqn:3}) and the equation of state parameter (\ref{eqn:33}), the energy conservation Eqns. (\ref{eqn:9},\ref{eqn:10}) can be written explicitly in the form 
\begin{eqnarray}
\dot{\rho_m}&=&\sqrt{3(\rho_m+\rho_d)}[b^2\rho_d-(1-b^2)\rho_m] \label{eqn:34}\\
\dot{\rho_d}&=&-\sqrt{3(\rho_m+\rho_d)}[b^2(\rho_m+\rho_d)+(1+\omega_d)\rho_d]. \label{eqn:35}
\end{eqnarray}
The above system is well defined and continuously differentiable on the open set $E=\lbrace (\rho_m,\rho_d)\in \mathbb{R}^2 | \rho_m>0, \rho_d>0\rbrace \subset \mathbb{R}^2$.  The critical points are $(\rho_m^c,\rho_d^c)\in E$ where $\rho_m^c=\frac{b^2}{(1-b^2)}\rho_d^c$ provided $1-b^2=\frac{\alpha}{\alpha-\beta}$ (we choose the cases in \cite{1309.3136[astro-ph.CO]} for which $\beta \leq 0$).  From Eqn. (\ref{eqn:33}) we derive $\omega_d=\frac{1}{b^2-1}(\leqslant -1)$.  At this set of critical points $\Omega_d=1-b^2$ and using (\ref{q2}) we get deceleration parameter $q=-1$.  The stability of the system (\ref{eqn:34},~\ref{eqn:35}) is identical to the system (\ref{eqn:23},~\ref{eqn:24}) by considering the same Liapunov function (\ref{eqn:25}) for exactly same set of critical points. 

\begin{center}
	B. $Q=\frac{\nu}{H}\rho_m\rho_d, ~(\nu>0)$
\end{center}
	The explicit form of the energy conservation equations are given in Eqns. (36) and (37) in reference \cite{1511.07955v1[gr-qc]}.  As a result, the evolution equation for $\Omega_d$ can be derived as follows
\begin{equation}\label{eqn:36}
\dot{\Omega_d}=-3H(1-\Omega_d)[(\alpha-1)-(\alpha-\beta)\Omega_d+\nu\Omega_d].
\end{equation}
This evolution equation for $\Omega_d$ together with the second Friedman equation 
\begin{equation}\label{eqn:37}
\dot{H}=-\frac{3H^2}{2}[\alpha-(\alpha-\beta)\Omega_d],
\end{equation}
form an autonomous system in $(\Omega_d,H)$ plane.\\
The critical points of the above autonomous system are $(\frac{\alpha}{\alpha-\beta}, H_c)$ for all $H_c \in \mathbb{R}-\lbrace 0 \rbrace$ provided $\nu= \frac{\alpha-\beta}{\alpha}>0$ (values of $\alpha$ and $\beta$ are defined in \cite{1309.3136[astro-ph.CO]}).  The eigenvalues of Jacobian matrix are $\lbrace (3H_c \beta(\frac{1}{\alpha}-1), 0\rbrace$ at the critical points.  Thus all critical points are non-hyperbolic critical points in nature.  To find the center manifold we shift the critical points to the origin.  The center manifold is $\Omega_d=0$ at the origin.  So the flow we get is $\dot{H}=0$ which actually reinforces the normally hyperbolic nature of the CPs.  Thus the nature of trajectory near each CP is determined by the vector fields along $\Omega_d$-axis and it is attracting to the CP if $3H_c \beta(\frac{1}{\alpha}-1)<0$ and repelling from the CP if $3H_c \beta(\frac{1}{\alpha}-1)>0$.  From Eqns. (\ref{eqn:33}) and (\ref{q2}) we obtain $\omega_d= \frac{\beta}{\alpha}-1$ and $q=-1$.\par
 The stability of CPs $(\Omega_d^c,0)$ (for $\Omega_d^c \in \mathbb{R}$) is not conclusive as there is no non-zero eigenvalue of the Jacobian matrix evaluated at these CPs.  Moreover, $\Omega_d$ is not defined for H=0.\par
  One may note that for a fixed positive (or negative) $H_c$ the system (\ref{eqn:36}-\ref{eqn:37}) is structurally unstable at $\beta(\frac{1}{\alpha}-1)=0$.  So each non-hyperbolic CP $(\frac{\alpha}{\alpha-\beta}, H_c)$ is bifurcation point at the bifurcation values $\beta=0$ and $\alpha=1$.

\begin{center}
	C. $Q=3\nu H\rho_d, ~(\nu>0)$
\end{center} 
In this case we consider the autonomous system in $(\Omega_d,H)$ plane as follows
\begin{eqnarray} 
\dot{\Omega_d}&=&-3H[(\alpha-1)(\Omega_d-1)^2+(1-\beta)\Omega_d(\Omega_d-1)+\nu \Omega_d] \label{eqn:38}\\
\dot{H}&=&-\frac{3H^2}{2}[\alpha-(\alpha-\beta)\Omega_d]. \label{eqn:39}
\end{eqnarray}
In this case we consider negative values of $\beta$ so that $\nu=-\frac{\beta}{\alpha}>0$ (as $\alpha>1$ in \cite{1309.3136[astro-ph.CO]}).  The critical points of system (\ref{eqn:38}-\ref{eqn:39}) are $(\frac{\alpha}{\alpha-\beta},H_c)$ for all $H_c \in \mathbb{R}-\lbrace 0 \rbrace$.  The eigenvalues of Jacobian matrix at each critical point are $\lbrace -3H_c(1+\beta + \nu),0\rbrace $.  The set of critical points are normally hyperbolic in nature.  So the critical points are stable if $-3H_c(1+\beta + \nu)<0$ and unstable if $-3H_c(1+\beta + \nu)>0$.  From Eqns. (\ref{eqn:33}) and (\ref{q2}) we obtain $\omega_d= \frac{\beta}{\alpha}-1$ and $q=-1$ which are identical to case B $(Q=\frac{\nu}{H}\rho_m\rho_d)$.\par
The stability of CPs $(\Omega_d^c,0)$ (for $\Omega_d^c \in \mathbb{R}$) is not conclusive as there is no non-zero eigenvalue of the Jacobian matrix evaluated at these CPs.  Moreover, $\Omega_d$ is not defined for H=0.\par
The system (\ref{eqn:38}-\ref{eqn:39}) is structurally unstable at the line $1+\beta +\nu=0$.  So each non-hyperbolic CP is a bifurcation point at the line of bifurcation values $\beta$ and $\nu$ satisfying $1+\beta+\nu=0$.

\section{Discussion and Cosmological implication}
The present work is an extensive study of dynamical system analysis of the autonomous systems formed by the cosmological evolution equations for different {\it IR}-cutoff in the context of Holographic Dark Energy (HDE) models.  In the present context the HDE is interacting with cold dark matter for three (popularly used in the literature) choices of the interaction term.  Here non-hyperbolic critical points are analyzed either by constructing the Liapunov function or by the center manifold theory. \par 
For the HDE model with event horizon as {\it IR}-cutoff we have line of critical points for all three choices of the interaction term.   The autonomous system formed by Eqns. (\ref{eqn:17}) and (\ref{eqn:18}) the line of critical points are characterized by $(H_c, \sqrt{3}CH_c)$, $H_c \in \mathbb{R}-\lbrace 0 \rbrace$ and $b^2=1-C^2$.  This line of critical points (CPs) are non-hyperbolic in nature as one of the eigenvalues of the Jacobian matrix is 0.  In particular, the CPs are normally hyperbolic and by analyzing the phase-space trajectories it is found that cosmologically there is a bouncing solution from expanding phase to contracting era of evolution.  Further, the above line of CPs are structurally unstable for $C=\pm \frac{1}{2}$ and hence the Universe experiences a bifurcation along the line of CPs with bifurcation values $C=\pm \frac{1}{2}$.  Also along the line of CPs $\Omega_d=C^2<1$ and $\omega_d=-\frac{1}{C^2}<-1$ and hence $q=-1$ with $\omega_T=-1$.  Thus the present cosmological model is in the phantom barrier although it is not fully dominated by HDE. \par 
The choice $b^2=1$ gives a simple form of autonomous system described by Eqns. (\ref{eqn:21}) and (\ref{eqn:22}) where the line of CPs $(H_c,0)$ still represents a bouncing cosmological solution.  This line of CPs are fully dominated by cold dark matter (with $q=\frac{1}{2}$) and the Universe is in a decelerating phase describing the dust era of evolution.  The matter evolution equations are converted into an autonomous system described by Eqns. (\ref{eqn:23}) and (\ref{eqn:24}).  Here the line of CPs $(\frac{b^2}{1-b^2}\rho_d^c, \rho_d^c)$, $\rho_d^c \in \mathbb{R}^{+}$ are as usual non-hyperbolic in nature.  By analyzing the Liapunov function (so constructed), it is found that the line of CPs are asymptotically  stable for $0\leqslant b^2 < \frac{1}{2}$ , stable at $b^2=\frac{1}{2}$ while they are unstable for $\frac{1}{2}< b^2 \leqslant 1$.  Hence $b=\pm \frac{1}{\sqrt{2}}$ are bifurcation values and corresponding bifurcation points is the line of CPs $(\rho_m^c,\rho_d^c)$.  Stability analysis of 3D-autonomous system (\ref{eqn:23},\ref{eqn:24},\ref{eqn:24b}) is same as system (\ref{eqn:23}-\ref{eqn:24}).\par 
For another choice of the interaction Eqns. (\ref{eqn:27}) and (\ref{eqn:28}) form an autonomous system having a plane of non-hyperbolic CPs $(u_c, H_c)$, $u_c,H_c \in \mathbb{R}-\lbrace 0 \rbrace$.  Here $\omega_d=-\frac{1}{\Omega_d}<-1$ and $q=-1$.  So HDE fluid behaves as phantom fluid.  However the effective single fluid behaves as cosmological constant (as $\omega_T=-1$).  For $C=1$ the system is structurally unstable at $\nu=1$ and hence the line of normally hyperbolic CPs $(1,H_c)$ are bifurcation points with bifurcation value $\nu=1$.  Another autonomous system is formed by the system of Eqns. (\ref{eqn:30}) and (\ref{eqn:31}) with the third choice of the interaction and the nature of the CPs are very similar to the above studies. \\
Finally HDE model has been studied with {\it IR}-cutoff at Ricci scalar curvature.  Three sets of autonomous systems are formed by the set of Eqns. $\lbrace (\ref{eqn:34}), (\ref{eqn:35}) \rbrace$, $\lbrace (\ref{eqn:36}),(\ref{eqn:37})\rbrace$ and $\lbrace (\ref{eqn:38}), (\ref{eqn:39}) \rbrace$ for three different choices of the interaction.  For the first autonomous system the line of CPs are similar to those in system of Eqns. (\ref{eqn:23}) and (\ref{eqn:24}) and the analysis is identical.  For the second set the line of CPs are $(\frac{\alpha}{\alpha-\beta}, H_c)$, $H_c \in \mathbb{R}-\lbrace 0 \rbrace$ with $\nu= \frac{\alpha-\beta}{\alpha}$.  By analyzing the phase-space trajectories near center manifold it is found that the system describes a bouncing solution for the cosmological evolution.  Also here the effective single fluid behaves as cosmological constant but the HDE is of phantom fluid in nature.  The line of CPs are bifurcation points at bifurcation values $\beta=0$ and $\alpha=1$.\par 
For the third set we have the same line of CPs with $\nu= -\frac{\beta}{\alpha}>0$ $(\beta<0)$.  The relevant cosmological parameters are $\omega_d=\frac{\beta}{\alpha}-1<1$, $\Omega_d=\frac{\alpha}{\alpha-\beta}$, $\omega_T=-1$, $q=-1$.  So cosmologically the evolution is very similar to the previous cases.  However, from the dynamical system point of view the system is structurally unstable along the parameter line $1+\beta+\nu=0$.  So line of CPs are bifurcation points having a line of bifurcation values $1+\beta+\nu=0$.  Therefore, from the above dynamical system analysis of the present interacting HDE cosmological model, we may conclude that the cosmic evolution is fully dominated by HDE having phantom fluid nature.  There is only one case where cold dark matter is the dominant part in the cosmic evolution and the Universe is in the dust era of evolution.  For future work, it will be interesting to give more physical interpretation in the cosmological context at the bifurcation points.  

\section{Acknowledgements}
The author S. Mishra is grateful to CSIR, Govt. of India for giving Junior Research Fellowship (CSIR File No: 09/096(0890)/2017- EMR - I) for the Ph.D work.  The authors are thankful to Dr. S. Pan of Presidency University for his valuable discussion.  SC thanks Science and Engineering Research Board (SERB) for awarding MATRICS
Research Grant support (File No: MTR/2017/000407).\\
Conflict of Interest: The authors declare that they have no conflict of interest.

\bibliography{Ref}
 
\end{document}